\documentclass{kapproc} 
\let\footnote\savefootnote
\let\footnotetext\savefootnotetext 
\let\footnoterule\savefootnoterule 
\kluwerbib

\RequirePackage{graphicx}%
\RequirePackage{epsf}%

\begin{document}

\articletitle{Chandra Detection of X-ray Absorption from Local Warm/Hot Gas}

\author{T. Fang$^{1,2}$, C. Canizares$^2$, K. Sembach$^3$, H. Marshall$^2$, J. Lee$^2$, D. Davis$^4$}
\affil{1. Dept. of Physics, Carnegie Mellon Univ., 5000 Forbes Ave., Pittsburgh, PA 15213\\
2. MIT, Center for Space Research, 70 Vassar St., Cambridge, MA 02139\\
3. STScI, 3700 San Martin Drive, Baltimore, MD  21218\\ 
4. GSFC, Code 661., GLAST SSC, Greenbelt, MD 20771}


\begin{abstract}
Recently, with the {\sl Chandra} X-ray Telescope we have detected several local X-ray absorption lines along lines-of-sight towards distant quasars. These absorption lines are produced by warm/hot gas located in local intergalactic space and/or in our Galaxy. I will present our observations and discuss the origin of the X-ray absorption and its implications in probing the warm/hot component of local baryons.
\end{abstract}

\section{Introduction}

The cosmic baryon budget at low and high redshift indicates that a
large fraction of baryons in the local universe have so far escaped
detection (e.g., Fukugita, Hogan, \& Peebles~1998). While there is
clear evidence that a significant fraction of these ``missing
baryons'' (between 20-40\% of total baryons) lie in photoionized,
low-redshift Ly$\alpha$ clouds (Penton, Shull, \& Stocke~2000), the
remainder could be located in intergalactic space with temperatures of
$10^{5}-10^{7}$ K (warm-hot intergalactic medium, or WHIM). Resonant
absorption from highly-ionized ions located in the WHIM gas has been
predicted based on both analytic studies of structure formation and
evolution (Shapiro \& Bahcall~1981;Perna \& Loeb~1998;Fang \& Canizares~2000) and cosmic
hydrodynamic simulations (Hellsten, Gnedin, \& Miralda-Escud{\'
e}~1998;Cen \& Ostriker~1999a;Dav{\' e} et al.~2001;Fang, Bryan, \&
Canizares~2002). Recent discovery of \ion{O}{6} absorption lines by
the Hubble Space Telescope ({\sl HST}) and the Far Ultraviolet
Spectroscopic Explorer ({\sl FUSE}) (see, e.g., Tripp \& Savage~2000) indicates that there may be
a significant reservoir of baryons in \ion{O}{6} absorbers. While
Li-like \ion{O}{6} probes about $\sim 30-40\%$ of the WHIM gas (Cen et
al.~2001;Fang \& Bryan~2001), the remaining $\sim 60-70\%$ is hotter
and can only be probed by ions with higher ionization potentials, such
as H- and He-like Oxygen, through X-ray observation.

Recently, with {\sl Chandra} Low Energy Transmission Grating Spectrometer (LETGS) we detected resonance absorption lines from H- and He-like Oxygen in the X-ray spectra of background quasars, namely PKS~2155-304 and 3C~273. The detected lines can be categorized into (1) those at $z \approx 0$ and (2) one redshifted intervening system. In this paper, we will discuss these detections and their implications for the physical properties of the hot gases that give rise to these absorption features.

\vbox{
\begin{center}
\begin{tabular}{cccc}
\multicolumn{4}{c}{Table 1: Fitting parameters of the X-ray absorption Lines~~~} \\
\hline
\hline
& \multicolumn{2}{c}{PKS~2155-304} & \multicolumn{1}{c}{3C~273} \\
\cline{2-3}\cline{4-4}
& O~{\sc viii} Ly$\alpha$ &  O~{\sc vii} He$\alpha$ & O~{\sc vii} He$\alpha$ \\
\hline
$\lambda_{obs}$       & $20.02_{-0.015}^{+0.015}$            & $21.61_{-0.01}^{+0.01}$       & $21.60_{-0.01}^{+0.01}$ \\ 
$cz({\rm km\ s^{-1}})$ & $16624\pm237$      & $112_{-138}^{+140}$  & $-26_{-140}^{+140}$ \\ 
Line Width$^{a}$      &  $< 0.039$                & $<0.027$             & $<0.020$\\ 
$\rm Line\ Flux^{b}$   & $4.8_{-1.9}^{+2.5}$       & $5.5_{-1.7}^{+3.0}$  & $4.2_{-0.9}^{+1.8}$ \\
EW (m${\rm \AA}$)      & $14.0_{-5.6}^{+7.3}$      & $15.6_{-4.9}^{+8.6}$ & $28.4_{-6.2}^{+12.5}$ \\ 
SNR              & 4.5                       & 4.6                  & 6.4 \\
\hline
\end{tabular}

\parbox{5in}{
\vspace{0.1in}
\small\baselineskip 9pt
\footnotesize
a. 90\% upper limit of the line width $\sigma$, in units of $\rm\AA$.\\
b. Absorbed line flux in units of $\rm 10^{-5}~photons~cm^{-2}s^{-1}$.\\
}
\end{center}
\normalsize
\centerline{}
}

\section{Data Reduction}

PKS~2155-304 and 3C~273 are bright extragalactic X-ray sources used as {\sl Chandra} calibration targets. They were observed
with the {\sl Chandra} LETG-ACIS (the observations ids for PKS~2155 are 1703, 2335, 3168; and the ids for 3C~273 are 1198, 2464, 2471). For detailed data analysis, we refer to Fang et al.~2002. We
found all continua are well described by a single power law absorbed by
Galactic neutral hydrogen.

After a blind search for any statistically significant absorption features, several absorption features with S/N~$>$~4 were detected in the spectra of both quasars in the 2--42~\AA\, region of the LETGS spectral bandpass (Figure~1). These
features were subsequently fit in ISIS (Houck \& Denicola~2000).

\section{Discussion}

\subsection{PKS~2155-304}

The absorption feature at $\sim 21.6$~\AA\, was reported by Nicastro
et al.~(2002) in the LETGS-HRC archival data. We concentrate on the
absorption feature which appears at $20.02~{\rm \AA}$ (619~eV).
Considering cosmic abundances and oscillator strengths  for different
ions, \ion{O}{8} Ly$\alpha$ is the only strong candidate line between
18 and 20~\AA\,,  the measured wavelength de-redshifted to the
source. It is plausible that the 20~\AA\ absorption is due to
\ion{O}{8} Ly$\alpha$ in a known intervening system at $cz\approx
16,734\rm\ km~s^{-1}$. With {\sl HST}, Shull et al.~(1998) discovered
a cluster of low metallicity \ion{H}{1} Ly$\alpha$ clouds along the
line-of-sight (LOS) towards PKS~2155-304, most of which  have
redshifts between $cz = 16,100\ {\rm km\ s^{-1}}$ and $18,500\ {\rm
km\ s^{-1}}$. Using 21 cm images from the {\sl Very Large Array} ({\sl
VLA}), they detected a small group of four \ion{H}{1} galaxies offset
by $\sim 400-800\ {\rm h_{70}^{-1}}$ kpc from the LOS, and suggested
that the \ion{H}{1} Ly$\alpha$ clouds could arise from gas associated
with the group (We use $\rm H_0 = 70h_{70}~km~s^{-1}Mpc^{-1}$
throughout the paper).

\begin{figure}[h]
\includegraphics[height=1.5in,width=5.5in]{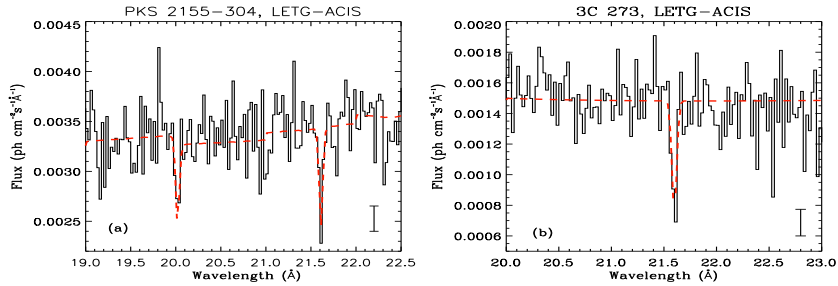}
\caption{The {\sl Chandra} LETG-ACIS spectra of (a) PKS~2155 and (b)
3C273. The dashed lines are the fitted spectra. The average
$1\sigma$ error bar plotted on the right-bottom of each panel is based
on statistics only.}
\end{figure}

Taking the absorption line to be \ion{O}{8} Ly$\alpha$, we estimate
the column density is $\rm N(O~VIII) \sim 9.5 \times 10^{15}~cm^{-2}$
if the line is unsaturated. We can constrain the density of the
absorbing gas, assuming it is associated with the intervening galaxy
group. Since the line is unresolved, a lower limit of $n_b>
(1.0\times10^{-5}\ {\rm cm}^{-3})\ Z_{0.1}^{-1}f_{0.5}^{-1}l_{8}^{-1}$
can be obtained. Here $Z_{0.1}$ is the metallicity in units of 0.1
solar abundance, $f_{0.5}$ is the ionization fraction in units of 0.5,
and $l_{8}$ is the path length in units of ${\rm 8h_{70}^{-1}Mpc}$. A
more reasonable estimate of the  path length comes from the mean
projected separation of $\sim$1 Mpc for the  galaxies in the group,
which gives $n_{b} \approx 7.5 \times 10^{-5}\ {\rm
cm}^{-3}~Z_{0.1}^{-1}f_{0.5}^{-1}$. This implies a range of baryon
overdensity ($\delta_{b} \sim$ 50 - 350) over the cosmic mean $\left<n_{b}\right> = 2.14\times
10^{-7}~\rm cm^{-3}$. Interestingly,
Shull et al.(1998) estimate an overdensity for the galaxy group of
$\delta_{gal}\sim$100.

In the case of pure collisional ionization, temperature is the only
parameter  of importance over a wide range of density so long as the
gas is optically thin.  The \ion{O}{8} ionization fraction peaks at
0.5, and exceeds 0.1 for temperatures $T \sim 2-5 \times10^{6}$
K. Using CLOUDY (Ferland et al.~1998) we find that photoionzation by
the cosmic UV/X-ray background is not important for $n_{b} > 10^{-5}\
{\rm cm^{-3}}$. CLOUDY calculations of the column density ratios
between other ions and \ion{O}{8} also show that $T \ga 10^{6.4}$ K.

Assuming  conservative upper limits of $\rm Z \la 0.5Z_{\odot}$ and an 
\ion{O}{8} ionization fraction of $f \la 0.5$ , and a path
length of $\Delta z \la 0.116$, we estimate $\rm \Omega_{b}(O~VIII)
\ga 0.005h_{70}^{-1}$. This is about 10\% of the total baryon
fraction, or about 30-40\% of the WHIM gas, if the WHIM gas contains
about 30-40\% of total baryonic matter. This baryon fraction is
consistent with the prediction from Perna \& Loeb~(1998) based on a
simple analytic model.

\subsection{3C~273}

Based on the detected line equivalent width ($W_{\lambda}$) and
non-detection of local \ion{O}{7} He$\beta$ line at $18.6288\AA$, we
estimate the column density $\rm N(O~{VII}) = 1.8_{-0.7}^{+0.2}
\times 10^{16}~cm^{-2}$ if the line is unsaturated\footnote{Rasmussen et al.(2002) also reported the detection of a similar feature with {\sl XMM}-Newton in this conference.}. The \ion{O}{7}
indicates the existence of the gas with high temperatures $\sim
10^{6}$ K. The extremely high temperatures imply that \ion{O}{7} is
unlikely to to be produced in the nearby interstellar medium (ISM),
although we cannot rule out the possibility of an origin from
supernova remnants. It is more plausible that the He-like Oxygen is
produced in distant, hot halo gas, or even in the Local Group (LG).

{\bf 3.2.1 Local Group Origin?}\\
We can constrain the density of the absorbing gas, assuming it is
associated with the Local Group. The path length can be set equal to
the distance to the boundary of the Local Group, where the gas begins
to participate in the Hubble flow. Assuming a simple
geometry for the Local Group, the path length is $\sim 1$ Mpc (see the
following text). Adopting the 90\% lower limit of the \ion{O}{7}
column density, this gives $n_{b} > (2.2\times10^{-5}\ {\rm cm}^{-3})\
Z_{0.2}^{-1}f_{1}^{-1}l_{1}^{-1}$, where $Z_{0.2}$ is the metallicity
in units of 0.2 solar abundance, $f_{1}$ is the ionization fraction, and $l_{1}$ is the path length in units of 1 Mpc.

Assuming spherical symmetry and isothermality, a model
characterizing the distribution of the Local Group gas is given by the
standard $\beta$-model. We adopt a simplified geometry model of the
LG, where the LG barycenter is located along the line connecting M31
and the Milky Way, at about 450 kpc away from our Galaxy (Rasmussen
\& Pedersen~2001). At about $r \sim 1200$ kpc the gravitational
contraction of the Local Group starts to dominate the Hubble flow,
and this was defined as the boundary of the Local Group (Courteau \&
van den Bergh~1999).  Based on this simple model we estimate the
column density of \ion{O}{7} by integrating the \ion{O}{7} number
density $n_{O~{VII}}$ along this path length. We find a tight upper
limit of Local Group temperature $\rm T \leq 1.2\times10^6$ K; at temperatures
higher than $1.2\times10^6$ K, the \ion{O}{7} ionization fraction drops quickly
and \ion{O}{8} starts to dominate. We also find that the temperature
of the Local Group should be higher than $2.3\times10^5$ K. To satisfy the
observed \ion{O}{7} column density we find that the gas distribution
should have a rather flat core with $r_{c} \geq 100$ kpc.

{\bf 3.2.2 Hot Halo Gas?}\\
Strong \ion{O}{6} absorption ($\rm \log N(O~VI) = 14.73\pm0.04$) along
the sight line towards 3C~273 was detected  with {\sl FUSE} between
-100 and +100 $\rm km\ s^{-1}$ (Sembach et al.~2001). This absorption
probably  occurs in the interstellar medium of the Milky Way disk and
halo. Absorption  features of lower ionization species are also
present at these velocities. \ion{O}{6} absorption is also detected
between +100 and +240 $\rm km\ s^{-1}$  in the form of a broad,
shallow absorption wing extending redward of the  primary  Galactic
absorption feature with $\rm \log N(O~VI) = 13.71$. The \ion{O}{6}
absorption wing  has been attributed to hot gas flowing out of the
Galactic disk as part of a "Galactic chimney" or "fountain" in the
Loop IV and North Polar Spur regions  of the sky. Alternatively, the
wing might be remnant tidal debris from  interactions of the Milky Way
and smaller Local Group galaxies (Sembach et al.~2001). It is possible
to associate the {\sl Chandra}-detected \ion{O}{7} absorption  with
these highly ionized metals detected by {\sl FUSE}. Here we discuss
several scenarios:

{\it (1). The \ion{O}{7} is related to the primary \ion{O}{6}
feature}: In this case, $\rm \log N(O~{VI}) = 14.73$, and  $\rm
\log [N(O~{VII})/N(O~{VI})]\sim1.5$, assuming $\rm N(O~{VII}) = 1.8\times10^{16}\ cm^{-2}$.  This is  within about a factor of
2 of the O~{\sc vii}/O~{\sc vi} ratio observed for the  PKS 2155-304
absorber and is consistent with the  idea that the gas is radiatively
cooling from a high temperature (Heckman et al.~2002).  This possibility is appealing since the  centroids of
the O~{\sc vi} and O~{\sc vii} absorption features are similar  ($\sim
6\pm10\ \rm km\ s^{-1}$  versus $-26\pm140\ \rm km\ s^{-1}$), and the
width of the resolved O~{\sc vi} line (FWHM $\sim$ 100 $\rm km\
s^{-1}$) is consistent with a broad O~{\sc vii} feature.  However,
this possibility also has drawbacks that the predicted O~{\sc viii}
column density is too high and the amount of C~{\sc iv} predicted is
too low.

{\it (2). The \ion{O}{7} is related to the \ion{O}{6} wing}: In this
case, the O~{\sc vii} is associated only with the  O~{\sc vi}
absorption "wing".  This seems like a reasonable possibility.   Then
$\rm \log [N(O~{VII})/N(O~{VI})]\sim2.5$. In collisional
ionization equilibrium, this would imply a temperature of $>10^6$ K
(Sutherland \& Dopita~1993). The non-detection of \ion{O}{8}
Ly$\alpha$ absorption requires the temperatures lower than
$\sim 10^{6.3}$ K.

{\it (3). The \ion{O}{7} is related to none of the \ion{O}{6}}: In
this case, the temperature should be high enough to
prevent the production of \ion{O}{6} ions. We can reach the similar conclusions to those in situation (2).

We thank members of the MIT/CXC team for their support. This work is supported in part by contracts NAS 8-38249 and SAO SV1-61010. KRS acknowledges financial support through NASA contract NAS5-32985 and Long Term Space Astrophysics grant NAG5-3485.



\begin{chapthebibliography}{1}
\bibitem[Cen \& Ostriker(1999a)]{cos99a} Cen, R. \& Ostriker, J.P. 1999a, ApJ, 514, 1
\bibitem[Cen \& Ostriker(1999b)]{cos99b} Cen, R.\ \& Ostriker, J.\ P.\ 1999b, ApJ, 519, L109
\bibitem[Cen et al.(2001)]{cto01} Cen, R., Tripp, T.M., Ostriker, J.P. \& Jenkins, E.B. 2001, ApJ, 559, L5
\bibitem[Courteau \& van den Bergh(1999)]{cva99} Courteau, 
S.~\& van den Bergh, S.\ 1999, AJ, 118, 337 
\bibitem[Dav{\' e} et al.(2001)]{dco01} Dav{\' e}, R.\ et al.\ 2001, ApJ, 552, 473
\bibitem[Fang \& Bryan(2001)]{fbr01} Fang, T.~\& Bryan, G.~L.\ 2001, ApJ, 561, L31
\bibitem[Fang, Bryan, \& Canizares(2002)]{fbc02} Fang, T., Bryan, G.L. \& Canizares, C.R. 2002, ApJ, 564, 604 
\bibitem[Fang \& Canizares(2000)]{fca00} Fang, T.~\& Canizares, C.~R.\ 2000, ApJ, 539, 532
\bibitem[Fang et al.(2002)]{fan02} Fang, T.~et al.\ 2002, ApJ, 572, L127 
\bibitem[Ferland et al.(1998)]{fkv98} Ferland, G.J.~et al.\ 1998, PASP, 110, 761
\bibitem[Fukugita, Hogan, \& Peebles(1998)]{fhp98} Fukugita, M., Hogan, C.~J., \& Peebles, P.~J.~E.\ 1998, ApJ, 503, 518. 
\bibitem[Houck \& Denicola(2000)]{hde00} Houck, J.~C.~\& Denicola, L.~A.\ 2000, ASP Conf.~Ser.~216: Astronomical Data Analysis 
Software and Systems IX, 9, 591 
\bibitem[Heckman, Norman, Strickland, \& Sembach(2002)]{hns02} Heckman, T.~M.~ et al~2002, ApJ, submitted (astro-ph/0205556)
\bibitem[Hellsten, Gnedin, \& Miralda-Escud{\' e}(1998)]{hgm98} Hellsten, U., Gnedin, N.~Y., \& Miralda-Escud{\' e}, J.\ 1998, ApJ, 509, 56
\bibitem[Nicastro et al.(2002)]{nic02} Nicastro, F.~et al.\ 
2002, ApJ, 573, 157
\bibitem[Penton, Shull, \& Stocke(2000)]{pss00} Penton, S.~V., Shull, J.~M., \& Stocke, J.~T.\ 2000, ApJ, 544, 150. 
\bibitem[Perna \& Loeb(1998)]{plo98} Perna, P. \& Loeb, A. 1998, ApJ,503, L135
\bibitem[Rasmussen \& Pedersen(2001)]{rpe01} Rasmussen, J.~\& Pedersen, K.\ 2001, ApJ, 559, 892 
\bibitem[Rasmussen~et al.(2002)]{ras02} Rasmussen, A.~et al.~2002, this proceeding
\bibitem[Sembach et al.(2001)]{sem01a} Sembach, K.~R.~et al.~2001, ApJ, 561, 573 
\bibitem[Shapiro \& Bahcall(1981)]{sba81} Shapiro, P.~R.~\& Bahcall, J.~N.\ 1981, ApJ, 245, 335
\bibitem[Shull et al.(1998)]{sps98} Shull, J.~M.~et al.\ 1998, AJ, 116, 2094
\bibitem[Sutherland \& Dopita(1993)]{sdo93} Sutherland, 
R.~S.~\& Dopita, M.~A.\ 1993, ApJS, 88, 253 
\bibitem[Tripp \& Savage(2000)]{tsa00} Tripp, T.M. \& Savage, B.D. 2000, ApJ, 542, 42

\end{chapthebibliography}
\end{document}